\documentclass[aps,10pt,prd,twocolumn,amsmath,amssymb,nofootinbib,superscriptaddress]{revtex4-1}
\usepackage[T1]{fontenc}
\usepackage{graphicx}
\usepackage{amsfonts}
\usepackage[colorlinks=true, linkcolor=blue, urlcolor=blue, citecolor=blue, anchorcolor=blue]{hyperref}
\usepackage{makecell, multirow}

\newcommand{\cA}{\ensuremath{\mathcal A} }
\newcommand{\cAbar}{\ensuremath{\overline{\mathcal A}} }
\newcommand{\Cbb}{\ensuremath{\mathbb C} }
\newcommand{\Ddag}{\ensuremath{D^{\dag}} }
\newcommand{\cD}{\ensuremath{\mathcal D} }
\newcommand{\cDbar}{\ensuremath{\overline{\mathcal D}} }
\newcommand{\cF}{\ensuremath{\mathcal F} }
\newcommand{\cFbar}{\ensuremath{\overline{\mathcal F}} }
\newcommand{\Ibb}{\ensuremath{\mathbb I} }
\newcommand{\cN}{\ensuremath{\mathcal N} }
\newcommand{\cO}{\ensuremath{\mathcal O} }
\newcommand{\cQ}{\ensuremath{\mathcal Q} }
\newcommand{\cU}{\ensuremath{\mathcal U} }
\newcommand{\cUbar}{\ensuremath{\overline{\mathcal U}} }
\newcommand{\al}{\ensuremath{\alpha} }
\newcommand{\be}{\ensuremath{\beta} }

\newcommand{\De}{\ensuremath{\Delta} }

\newcommand{\la}{\ensuremath{\lambda} }
\newcommand{\lalat}{\ensuremath{\la_{\text{lat}}} }
\newcommand{\muhat}{\ensuremath{\widehat\mu} }

\newcommand{\SO}[1]{\ensuremath{\text{SO(}#1\text{)}} }
\newcommand{\X}{\ensuremath{\!\times\!} }

\newcommand{\Tr}[1]{\ensuremath{\text{Tr}\left[ #1 \right]} }
\newcommand{\vev}[1]{\ensuremath{\left\langle #1 \right\rangle} }
\newcommand{\eq}[1]{Eq.~(\ref{#1})}
\newcommand{\fig}[1]{Fig.~\ref{#1}}
\newcommand{\tab}[1]{Table~\ref{#1}}
\newcommand{\secref}[1]{Sec.~\ref{#1}}
\newcommand{\refcite}[1]{Ref.~\cite{#1}}

\newcommand{\Uone}{\ensuremath{\text{U(1)}} }

\begin{document}
\title{Nonperturbative phase diagram of two-dimensional ${\cal N} = (2, 2)$ super-Yang--Mills} 

\author{Navdeep~Singh~Dhindsa}\email{navdeep.s.dhindsa@gmail.com}
\affiliation{Department of Physical Sciences, Indian Institute of Science Education and Research - Mohali, Knowledge City, Sector 81, SAS Nagar, Punjab 140306, India}
\affiliation{\mbox{The Institute of Mathematical Sciences, a CI of Homi Bhabha National Institute}, Chennai, 600113, India}

\author{Raghav~G.~Jha}\email{raghav.govind.jha@gmail.com}
\affiliation{Thomas Jefferson National Accelerator Facility, Newport News, VA 23606, United States}

\author{Anosh~Joseph}\email{anosh.joseph@wits.ac.za}
\affiliation{National Institute for Theoretical and Computational Sciences, School of Physics, and Mandelstam Institute for Theoretical Physics, University of the Witwatersrand, Johannesburg, Wits 2050, South Africa}

\author{David~Schaich}\email{david.schaich@liverpool.ac.uk}
\affiliation{Department of Mathematical Sciences, University of Liverpool, Liverpool L69 7ZL, United Kingdom}

\begin{abstract} 
We consider two-dimensional ${\cal N} = (2, 2)$ Yang--Mills theory with gauge group SU($N$) in Euclidean signature compactified on a torus with thermal fermion boundary conditions imposed on one cycle.  
We perform non-perturbative lattice analyses of this theory for large $12 \leq N \leq 20$.
Although no holographic dual of this theory is yet known, we conduct numerical investigations to check for features similar to the two-dimensional ${\cal N} = (8, 8)$ Yang--Mills theory, which has a well-defined gravity dual.
We perform lattice field theory calculations to determine the phase diagram, observing a spatial deconfinement transition similar to the maximally supersymmetric case.
However, the transition does not continue to low temperature, implying the absence of a topology-changing transition between black hole geometries in any holographic dual for this four-supercharge theory.
\end{abstract}

\maketitle

\section{Introduction}
\label{sec:intro}

A version of the holographic duality conjecture relates weakly coupled gravitational theories in $D+1$ spacetime dimensions to strongly coupled SU($N$) supersymmetric Yang--Mills (SYM) theories with maximal supersymmetry in $D$ spacetime dimensions in the large-$N$ limit~\cite{Maldacena:1997re, Itzhaki:1998dd}. Studying strongly coupled pure SYM theories using analytical methods can be extremely hard. A lattice formulation of these theories provides an inherently non-perturbative way to investigate them at strong coupling and finite $N$. A naive lattice regularization of a supersymmetric field theory would explicitly break supersymmetry. However, for certain classes of SYM theories, it is possible to preserve a subset of the supersymmetries at non-zero lattice spacing. In particular, these supersymmetric lattice constructions require at least $2^D$ supercharges in $D$ spacetime dimensions. This condition is satisfied for pure SYM theories with sixteen supercharges, with known large-$N$ holographic duals, for all $2 \leq D \le 4$. See Ref.~\cite{Catterall:2009it} for a thorough review of these constructions and Ref.~\cite{Schaich:2022xgy} for a review of more recent work.

The two-dimensional $\cN = (8, 8)$ SU($N$) SYM theory at finite temperature and compactified on a spatial circle has a well-defined holographic dual at large $N$. The dual gravitational system, at low temperature, contains various types of black hole solutions, with compact spatial circle, arising in Type IIA and IIB supergravity. The phase diagram of the gravitational system is expected to contain a region where homogeneous D1 (black string) solutions that wrap around the spatial circle dominate and another in which D0 (black hole) solutions that are localized on the spatial circle dominate. At low temperature, there exists a first-order Gregory--Laflamme (GL) phase transition between the homogeneous and localized solutions \cite{Gregory:1993vy, Li:1998jy, Martinec:1998ja, Kol:2002xz, Aharony:2004ig, Harmark:2004ws, Aharony:2005ew}.
In the dual gauge theory, this corresponds to a `spatial deconfinement' transition with the spatial Wilson line magnitude serving as an order parameter. 

In this work, we focus on the gauge theory system with a lower number of supercharges: two-dimensional $\cN = (2, 2)$ SU($N$) SYM theory on a Euclidean torus. Although a well-defined holographic dual has not yet been constructed for this theory, we aim to understand how much it resembles its sixteen-supercharge counterpart, which has a well-defined gravity dual. It is also interesting to consider how the reduction of supersymmetry can affect the holographic features of a given theory. We use lattice field theory to investigate the phase structure of $\cN = (2, 2)$ SYM and compare this to the corresponding lattice results for $\cN = (8, 8)$ SYM~\cite{Catterall:2010fx, Catterall:2017lub}. Specifically, we map out the line of spatial deconfinement transitions as a function of the shape of the torus quantified by its aspect ratio [\eq{eq:aspect_ratio}].

There have been several prior lattice studies of the two-dimensional $\cN = (2, 2)$ SYM theory, including Refs.~\cite{Catterall:2006jw, Suzuki:2007jt, Kanamori:2007ye, Kanamori:2007yx, Kanamori:2008bk, Catterall:2008dv, Kanamori:2008yy, Kanamori:2009dk, Hanada:2009hq, Hanada:2010qg, Catterall:2011aa, Mehta:2011ud, Galvez:2011ief, Kamata:2016xmu, Catterall:2017xox, August:2018esp, Dhindsa:2021irw}.
In particular, Refs.~\cite{Hanada:2010qg, Catterall:2011aa, Mehta:2011ud, Galvez:2011ief, Catterall:2017xox} have established that the theory does not exhibit a sign problem for sufficiently small lattice spacings.
In this work, we will ensure that our numerical calculations remain in this sign-problem-free regime.

The paper is organized as follows: In \secref{Sec:2d_Q4}, we present the supersymmetric lattice construction of the two-dimensional ${\cal N} = (2, 2)$ SYM theory at finite-temperature on a two-torus.
In \secref{sec:Lattice_calcs}, we discuss our results for the spatial deconfinement transition and the behavior of the critical temperature for aspect ratios $1/2 \leq \al \leq 4$.
The data leading to these results are available through~\refcite{data}.
Finally, we conclude in \secref{Sec:Conclusions} and discuss promising directions for future work.

\section{Two-dimensional ${\cal N} = (2, 2)$ lattice SYM theory}
\label{Sec:2d_Q4}

The two-dimensional $\cN = (2, 2)$ SYM theory can be obtained by dimensionally reducing four-dimensional $\cN = 1$ SYM. The parent theory, in Euclidean spacetime, has a global symmetry group $\SO{4}_E \times \Uone$, with $\SO{4}_E$ and \Uone denoting the Euclidean rotation symmetry and chiral symmetry, respectively. After dimensional reduction, the global symmetry group becomes $\SO{2}_E \times \SO{2}_{R_1} \times \Uone_{R_2}$, with $\SO{2}_E$ denoting the Euclidean rotation symmetry, $\SO{2}_{R_1}$ the rotation symmetry along the reduced dimensions, and $\Uone_{R_2}$ the chiral symmetry.

We turn to a topologically twisted version of the original two-dimensional theory to construct the lattice $\cN = (2, 2)$ SYM theory that preserves one of the four supercharges. In flat Euclidean spacetime, twisting amounts to relabelling the fields and supercharges of the original theory. We can take the diagonal sub-group of the rotation groups, $\SO{2}_E$ and $\SO{2}_{R_1}$, to define
\begin{equation}
  \SO{2}' \equiv \text{diag} \Big(\SO{2}_E \times \SO{2}_{R_1}\Big),
\end{equation}
which we call the twisted rotation group. After twisting, the fields and supercharges of the untwisted theory rearrange themselves to form representations of the twisted rotation group. The four supercharges of the original theory combine in a way that they become integer-spin supercharges of the twisted theory: a scalar $\cQ$, a vector $\cQ_a$ with $a = 1$, $2$, and an antisymmetric tensor $\cQ_{ab} = -\cQ_{ba}$. The fermionic degrees of freedom rearrange themselves into a set of geometric fermions comprising of a scalar fermion $\eta$, a vector fermion $\psi_a$ and an antisymmetric tensor fermion $\chi_{ab}$. Rearrangements of the fields happen in the bosonic sector as well --- the gauge field $A_a$ and the two scalars $X_a$ combine to form a complexified gauge field $\cA_a = A_a + i X_a$ that transforms as a vector under the twisted rotation group.\footnote{This complexification promotes SU($N$) to U($N$) gauge invariance, as discussed by \refcite{Schaich:2022xgy} and references therein.  For notational convenience, we continue to refer to SU($N$) gauge groups.}

After twisting, the action of the original theory can be expressed in a $\cQ$-exact form: $S = \cQ \Psi$, with
\begin{equation}
  \Psi = \frac{N}{4 \la} \int d^2x ~\text{Tr} \left(\chi_{ab} \cF_{ab} + \eta \left[ \overline{\cD}_a, \cD_a \right] - \frac{1}{2} \eta d \right),
\end{equation}
summing over repeated indices.
Here \la is the 't~Hooft coupling, $\cF_{ab} = \left[\cD_a, \cD_b\right]$ is the complexified field strength, and $\cD_a = \partial_a + \cA_a$ is the complexified covariant derivative. We use an anti-Hermitian basis for the generators of the gauge group with $\text{Tr}(T_A T_B) = - \delta_{AB}$.
The scalar supercharge $\cQ$ acts on the twisted fields in the following (nilpotent) way:
\begin{align*}
    \cQ \cA_a & = \psi_a, &
    \cQ \psi_a & = 0, \\
    \cQ \chi_{ab} & = - \cFbar_{ab}, &
    \cQ \cAbar_a & = 0, \\
    \cQ \eta & = d, &
    \cQ d & = 0.
\end{align*}
Here $\cFbar_{ab} = \left[\cDbar_a, \cDbar_b\right]$ with $\cDbar_a = \partial_a + \cAbar_a$ and $\cAbar_a = A_a - i X_a$.
We have also introduced a bosonic auxiliary field $d$ for the off-shell completion of the supersymmetry algebra.
It obeys the equation of motion 
\begin{equation}
  \label{eq:EOM}
  d = \left[\cDbar_a, \cD_a\right].
\end{equation}
After performing the \cQ variation on $\Psi$ and integrating over the auxiliary field $d$, the continuum twisted action takes the form
\begin{equation}
  \label{eq:twisted_action_continuum}
  \begin{split}
    S = \frac{N}{4 \la} \int d^2x~\text{Tr} \Bigg( & - \cFbar_{ab} \cF_{ab} + \frac{1}{2} \left[\cDbar_a, \cD_a \right]^2 \\
                                                   & - \chi_{ab} \cD_{[a} \psi_{b]} - \eta \cDbar_a \psi_a \Bigg).
  \end{split}
\end{equation}
It is easy to show that this action is invariant under the scalar supercharge: $\cQ S = 0$.
Using the equations of motion, one can show that it is also invariant under the remaining three twisted supercharges~\cite{Catterall:2006jw, Catterall:2009it, Catterall:2013roa}. 

We can construct a supersymmetric lattice action from the twisted action given in \eq{eq:twisted_action_continuum}.
The lattice action remains exactly invariant under the scalar supercharge $\cQ$, thanks to the fact that this is still a nilpotent supercharge, $\cQ^2 = 0$, at non-zero lattice spacing.
Note that the other three twisted supercharges, though nilpotent in the continuum, are no longer nilpotent on the lattice, and thus they are broken at non-zero lattice spacing.
However, it can be shown that we recover these broken supersymmetries, along with the Euclidean rotational symmetry, as the continuum limit of the lattice theory is taken~\cite{Cohen:2003xe, Catterall:2006jw, Catterall:2009it, Catterall:2013roa}.

We discretize the twisted action on a two-dimensional square lattice spanned by two orthogonal unit vectors, $\muhat_a$, with $a = 1$, $2$. We employ the geometrical discretization scheme provided in Ref.~\cite{Damgaard:2008pa} to place the fields and their derivatives on the lattice unit cell in a gauge-invariant way. 

On the lattice, the complexified gauge field is mapped to a complexified link, $\cA_a(x) \rightarrow \cU_a(n)$, which is an element of the algebra $\mathfrak{gl}(N, \Cbb)$.
The field $\cU_a(n)$ lives on an oriented link connecting sites $n$ and $n + \muhat_a$.
Its superpartner, the vector fermion $\psi_a(n)$, is also a link variable oriented along the same link.
Similarly, $\eta(n)$ lives on the site $n$ while $\chi_{ab}(n)$ lives on the diagonal of the unit cell. The placements and orientations of the fields ensure a gauge-invariant lattice action.
Finite difference operators replace the covariant derivatives according to the rules given in \refcite{Damgaard:2008pa}. We have
\begin{equation}
  \begin{split}
    \cDbar_a^{(-)} f_a(n) &= f_a(n) \cUbar_a(n) - \cUbar_a(n - \muhat_a) f_a(n - \muhat_a), \\
    \cD_a^{(+)} f_b(n) &= \cU_a(n) f_b(n + \muhat_a) - f_b(n) \cU_a(n + \muhat_b).
  \end{split}
\end{equation}
The lattice action, which is local, doubler-free, gauge invariant, and $\cQ$ supersymmetric, is then
\begin{align}
  S = \frac{N}{4 \lalat} & \sum_n \text{Tr} \Bigg[ - \cFbar_{ab}(n) \cF_{ab}(n) + \frac{1}{2} \left(\cDbar_a^{(-)} \cU_a(n)\right)^2 \nonumber \\
                         & - \chi_{ab}(n) \cD^{(+)}_{~[a}\psi^{\ }_{~b]}(n) - \eta(n) \overline{\cD}_a^{(-)} \psi_a(n) \Bigg]. \label{eq:exact}
\end{align}
In addition to this action, we add a scalar potential with a tunable coefficient $\mu$ to lift flat directions in the theory.
While this softly breaks the scalar \cQ supersymmetry, it is needed in order to carry out numerical calculations. Thus, the complete action is given by
\begin{equation}
  \label{eq:soft}
  S_{\text{total}} = S + \frac{N \mu^2}{4 \lalat}\sum_{n, a} \Tr{\left(\cUbar_a(n) \cU_a(n) - \Ibb_N \right)^2}.
\end{equation}

\section{Lattice calculations}
\label{sec:Lattice_calcs}

\subsection{Lattice setup}
\label{sec:Lattice_setup}
We denote as $N_\tau$ and $N_x$ the number of lattice sites along the temporal and spatial directions.
With \be and $L$ the dimensionful temporal and spatial extents, respectively, we have
\begin{align}
  \be & = a N_\tau, &
  L & = a N_x.
\end{align}
We obtain a torus by imposing thermal boundary conditions (BCs), which are periodic for all fields along the spatial direction and periodic (anti-periodic) for bosons (fermions) along the temporal direction.
All fields and variables on the lattice are made dimensionless using the dimensionful 't~Hooft coupling $\la$.
We define dimensionless temporal and spatial extents as
\begin{subequations}
\begin{align}
  r_\tau & \equiv \be \sqrt{\la} = N_\tau \sqrt{\lalat} = 1/t, \\
  r_x & \equiv L \sqrt{\la} = N_x \sqrt{\lalat},
\end{align}
\end{subequations}
with $t$ denoting the dimensionless temperature and $\lalat \equiv \lambda a^2$ the dimensionless 't~Hooft coupling.
We also define the aspect ratio \al mentioned above, which can be expressed in any of the three ways:
\begin{equation}
  \label{eq:aspect_ratio}
  \al = \frac{L}{\be} = \frac{r_x}{r_\tau} = \frac{N_x}{N_\tau}.
\end{equation}
We carry out numerical calculations with fixed $r_x$ and $r_{\tau}$, meaning that the $a \to 0$ continuum limit corresponds to $N_{x, \tau} \to \infty$ while $\lalat \to 0$.

To ensure that the supersymmetry-breaking scalar potential we added to the action in \eq{eq:soft} is automatically removed in the $\lalat \to 0$ continuum limit, we set
\begin{equation}
  \label{eq:zeta}
  \mu = \zeta \frac{r_\tau}{N_\tau} = \zeta \sqrt{\lambda} a = \zeta \sqrt{\lalat},
\end{equation}
and carry out numerical computations with fixed (dimensionless) $\zeta$.

This two-dimensional supersymmetric lattice system is implemented in the publicly available parallel software~\cite{susy_code} presented by \refcite{Schaich:2014pda}.
This package provides a rational hybrid Monte Carlo (RHMC) algorithm and measurements of the Wilson lines, scalar eigenvalues, extremal eigenvalues and pfaffian phase of the fermion operator, and other observables of interest.
Using this code, numerical results for the $\cN = (2, 2)$ SYM vacuum energy were presented in \refcite{Catterall:2017xox}.
There, it was shown that the vacuum energy density vanishes within uncertainties, which is consistent with the absence of dynamical supersymmetry breaking in this theory.
\refcite{Catterall:2017xox} also investigated the phase of the pfaffian for $1 \leq r_{\tau} \leq 9$, finding that the phase fluctuations are negligible throughout this range and vanish as the continuum limit is taken.

In our numerical calculations, it is necessary to choose values of $\zeta$ that are large enough to lift the flat directions without introducing excessive soft \cQ supersymmetry breaking.
We confirm the successful lifting of flat directions by monitoring the Maldacena loop
\begin{equation}
  \label{eq:Maldacena}
  M \equiv \frac{1}{N N_x} \sum_{x = 0}^{N_x - 1} \Tr{\prod_{t = 0}^{N_{\tau} - 1} \cU_t(x, t)},
\end{equation}
and ensuring that it is stable with an $\cO(1)$ magnitude.
The largest value is $\vev{|M|} = 3.97(9)$, for an ensemble with gauge group SU(12), lattice size $12\X 12$, and a high temperature corresponding to $r_{\tau} = 0.6$. 
Full numerical results are available in the open data release \refcite{data}, along with similar sanity checks of the RHMC acceptance rate and the Creutz equality $\vev{e^{-\De H}} = 1$~\cite{Creutz:1988wv}.

Similarly, we monitor violations of two \cQ Ward identities to ensure that $\zeta$ is not too large.
As discussed in more detail by \refcite{Catterall:2015ira}, each Ward identity can be expressed as the vacuum expectation value of the supersymmetry transformation of a suitable local operator, $\vev{\cQ \cO}$.
The lattice action in \eq{eq:exact} already provides one such local operator.
Because the fermion action is gaussian, this Ward identity fixes the bosonic action per lattice site to be $s_B = 3N^2/ 2$.
We can therefore use
\begin{equation*}
  \frac{\left|\vev{s_B} - 1.5N^2\right|}{1.5N^2}
\end{equation*}
as a normalized measure of its violation.
The largest violation is 0.0234(3) for an ensemble with gauge group SU(12), lattice size $12\X 12$, and a low temperature corresponding to $r_{\tau} = 5$. 

Similarly, \refcite{Catterall:2014vka} pointed out that another suitable Ward identity is provided by
\begin{equation}
  \label{eq:bilin}
  \cQ \Tr{\eta \cU_a \cUbar_a} = \Tr{\cDbar_b^{(-)}\cU_b \cU_a \cUbar_a} - \Tr{\eta \psi_a \cUbar_a},
\end{equation}
using the equation of motion for the bosonic auxiliary field, \eq{eq:EOM}.
It is convenient to introduce the shorthand notation $B \equiv \Tr{\cDbar_b^{(-)}\cU_b \cU_a \cUbar_a}$ for the purely bosonic term in this difference, and $F \equiv \Tr{\eta \psi_a \cUbar_a}$ for the term involving the $\eta\psi$ fermion bilinear that we compute stochastically using random gaussian noise vectors.
Then
\begin{equation*}
  \left|\vev{\frac{B - F}{\sqrt{B^2 + F^2}}}\right|
\end{equation*}
provides a normalized measure of the violation of this Ward identity.
Its largest value is 0.0375(7), again for gauge group SU(12), lattice size $12\X 12$, and $r_{\tau} = 5$. 
All results for these two Ward identities are also available in the open data release \refcite{data}.

\subsection{Spatial deconfinement transition}

In this work, we focus on the behavior of the Wilson lines wrapping around the torus in either the spatial or temporal direction for various values of $r_\tau$ and $r_x$.
As discussed above, for the maximally supersymmetric $\cN = (8, 8)$ SYM theory, the spatial deconfinement phase transition signaled by the spatial Wilson line is related to a topology-changing transition between black-string and black-hole geometries in the holographic dual supergravity solutions.
This transition was observed through lattice field theory calculations in Refs.~\cite{Catterall:2010fx, Catterall:2017lub}.
That work also monitored the Polyakov loop (that is, the temporal Wilson line) to ensure the system was thermally deconfined, which is needed for holographic duality to hold in principle~\cite{Witten:1998zw}.
We now present our new numerical results for two-dimensional $\cN = (2, 2)$ SYM theory, which indicate that the spatial deconfinement transition in this theory is restricted to the high-temperature regime $r_\tau \lesssim 1$.

%
\begin{figure*}[htbp]
  \includegraphics[width=0.49\linewidth]{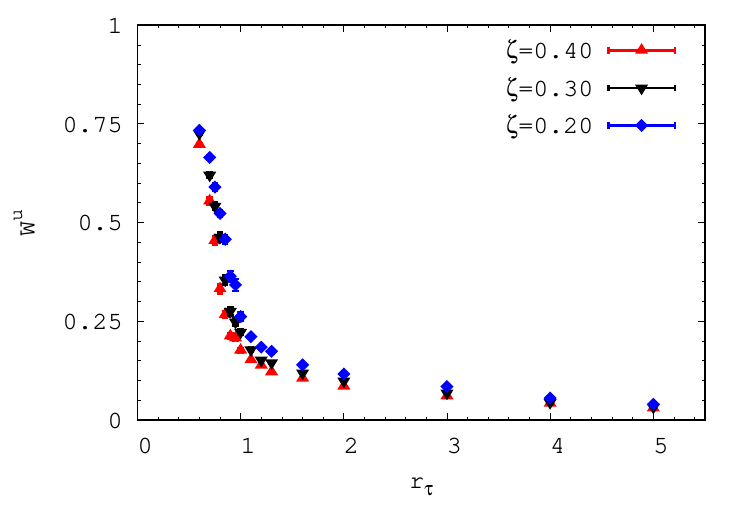} \hfill
  \includegraphics[width=0.49\linewidth]{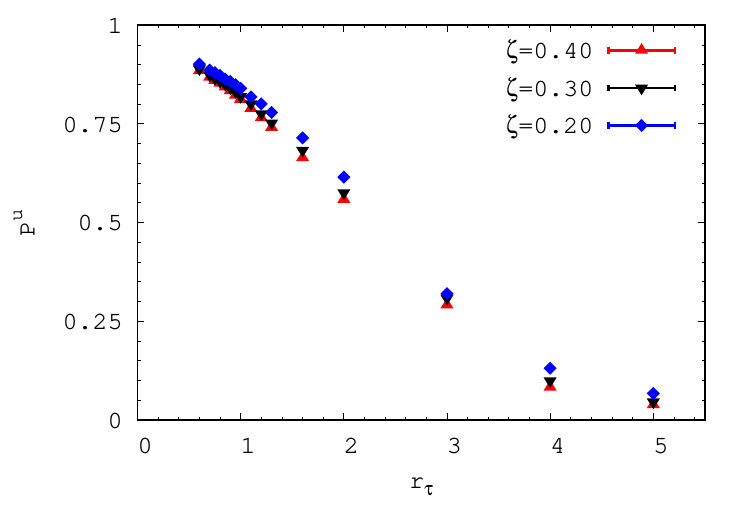}
  \caption{The $r_\tau$ dependence of the magnitudes of the unitarized spatial Wilson line $\vev{|W^u|}$ (left) and Polyakov loop $\vev{|P^u|}$ (right) on $12\X 12$ lattices with gauge group SU(12).}
  \label{Fig:wilson}
\end{figure*}

\begin{figure}[htbp]
  \includegraphics[width=\linewidth]{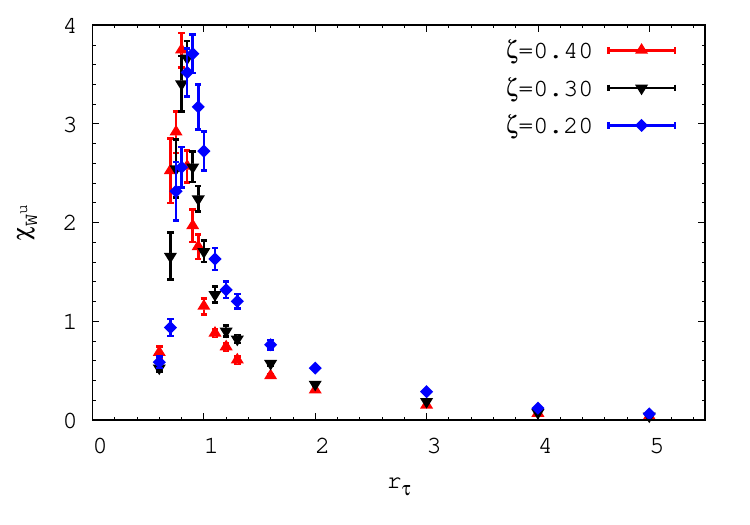}
  \caption{The $r_\tau$ dependence of the susceptibility of $\vev{|W^u|}$, for the same SU(12) $12\X 12$ ensembles considered in \fig{Fig:wilson}.}
  \label{Fig:susceptibility_spatial_Wilson}
\end{figure}

The specific observable we use to analyze the phase transition is the unitarized spatial Wilson line,
\begin{equation}
  \label{eqn:wilson_spatial}
  W^u \equiv \frac{1}{N N_{\tau}} \sum_{t = 0}^{N_{\tau} - 1} \Tr{\prod_{x = 0}^{N_x - 1} U_x(x, t)}.
\end{equation}
Here, $U_a(n)$ is extracted from the polar decomposition
\begin{equation}
  \cU_a(n) = e^{X_a(n)}U_a(n),
\end{equation}
where we can identify $U_a(n)$ as the unitarized gauge link connecting sites $n$ and $n + \muhat_a$, and $X_a(n)$ as the scalar field at site $n$.
In the spatially deconfined phase, $W^u$ sits in one of the degenerate $Z_N$ vacua with a large magnitude, spontaneously breaking the $Z_N$ center symmetry.
Note that \eq{eqn:wilson_spatial} normalizes $|W^u| \leq 1$, with equality for the free theory, so ``large'' in this context means $0.5 \lesssim \vev{|W^u|} \leq 1$.
In the spatially confined phase, $\vev{|W^u|} \to 0$ in the large-$N$ limit.

We also monitor the unitarized Polyakov loop
\begin{equation}
  P^u \equiv \frac{1}{N N_x} \sum_{x = 0}^{N_x - 1} \Tr{\prod_{t = 0}^{N_{\tau} - 1} U_t(x, t)}.
\label{eqn:wilson_temporal}
\end{equation} 
The calculations need to remain in the thermally deconfined phase corresponding to a large Polyakov loop magnitude, $0.5 \lesssim \vev{|P^u|} \leq 1$, to admit, in principle, a holographic dual interpretation in terms of a black hole geometry.

In \fig{Fig:wilson}, we show the $r_\tau$ dependence of $\vev{|W^u|}$ and $\vev{|P^u|}$ from a subset of our lattice ensembles, with gauge group SU(12) and $12\X 12$ lattice size.
See Table~\ref{Tab:config} in Appendix~\ref{Sec:Lattice_ensembles} for a brief summary of our numerical calculations.
For this aspect ratio $\al = 1$, we have $r_x = r_{\tau}$.
We see that $\vev{|W^u|}$ indicates a spatial deconfinement transition around $r_\tau \approx 0.85$.
The results for $\vev{|P^u|}$ confirm that the system remains thermally deconfined in this regime and up to $r_\tau \sim 2$.

To analyze the spatial deconfinement transition in more detail, we compute the susceptibility of the spatial Wilson line, defined as
\begin{equation}
  \chi_{_{W^u}} \equiv N^2 \left(\vev{|W^u|^2} - \vev{|W^u|}^2 \right).
  \label{eqn:susc}
\end{equation}
Figure~\ref{Fig:susceptibility_spatial_Wilson} shows the susceptibility for the same ensembles.
A clear peak is visible around the $r_\tau \approx 0.85$ expected from \fig{Fig:wilson}.
The location of the susceptibility peak provides the critical inverse temperature $r_{\tau}^{(c)}$ of the spatial deconfinement transition for this gauge group and lattice size.
While these $r_{\tau}^{(c)}$ shift slightly to lower temperatures as we reduce the value of $\zeta$ used to lift the flat directions [\eq{eq:zeta}], within uncertainties their $\zeta \to 0$ extrapolation is consistent with a constant.

In addition, by analyzing how the height of this peak $\chi_{\text{max}}$ changes with the number of degrees of freedom, we can estimate the order of the transition.
By fixing the lattice size so that the number of degrees of freedom scales $\propto N^2$ for different SU($N$) gauge groups, we can fit the $N$ dependence of $\chi_{\text{max}}$ as
\begin{equation}
  \label{eq:power}
  \chi_{\text{max}} = C N^{2b}
\end{equation}
with $C$ and $b$ as fitting parameters.
In the case of a first-order transition, we expect the peak height to scale proportionally to the number of degrees of freedom, $\chi_{\text{max}} \propto N^2$, that is, with $b = 1$.
For a crossover, the peak height is independent of $N$ (i.e., $b = 0$), while a continuous second-order transition is characterized by $0 < b < 1$ that corresponds to a critical exponent of the theory~\cite{Imry:1980zz, Fisher:1982xt, Binder:1984llk, Challa:1986sk, Fukugita:1989yb}.

\begin{figure*}[htbp]
  \includegraphics[width=0.49\linewidth]{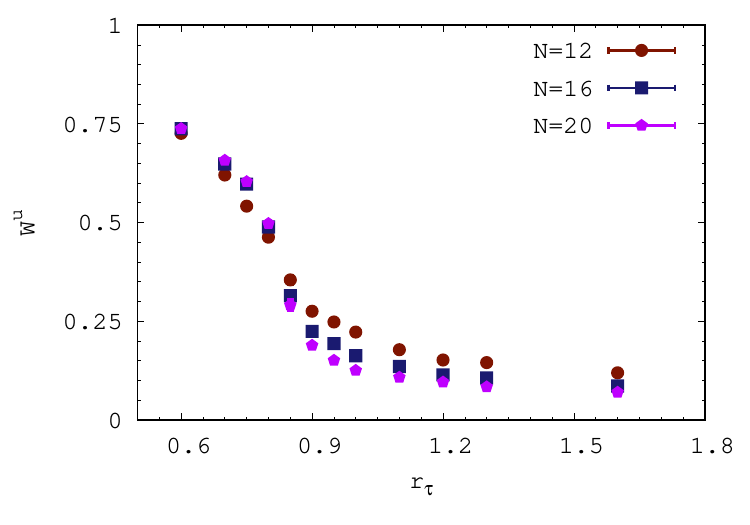} \hfill
  \includegraphics[width=0.49\linewidth]{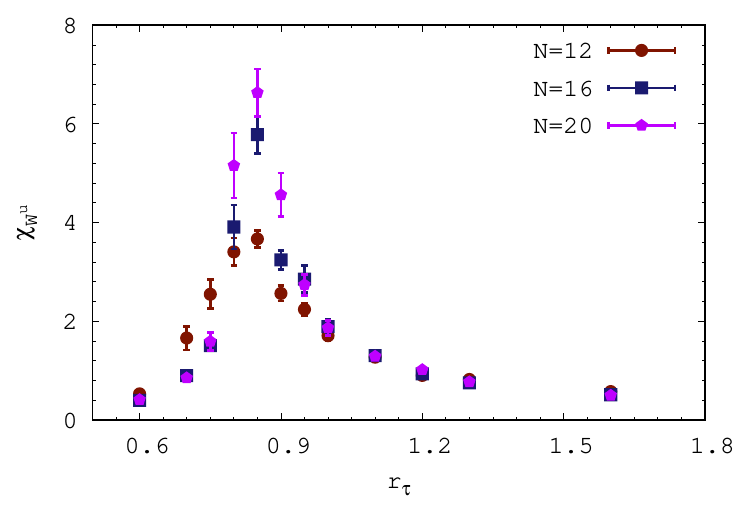}
\caption{The spatial Wilson line (left) and its susceptibility (right) for gauge groups SU(12), SU(16) and SU(20), all from $12\X 12$ lattices with $\zeta = 0.3$.}
\label{Fig:chi_N}
\end{figure*}

\begin{figure}[htbp]
  \includegraphics[width=0.9\linewidth]{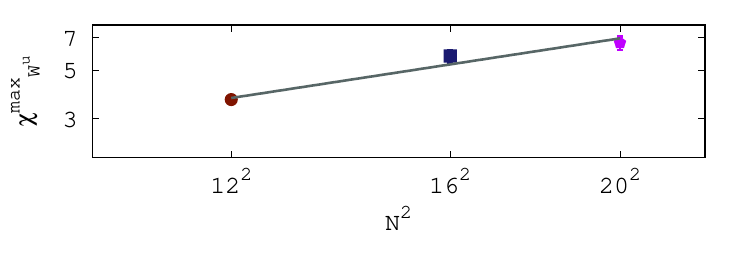}
  \caption{The height of the susceptibility peaks in \fig{Fig:chi_N}, $\chi_{\text{max}}$, vs.\ $N^2$ on log--log axes, including a power-law fit.}
  \label{Fig:scaling}
\end{figure}
    
To estimate the order of the transition, we carried out additional $12\X 12$ lattice calculations with $N = 16$ and $20$.
In \fig{Fig:chi_N} we present our results for $\vev{|W^u|}$ and $\chi_{_{W^u}}$ vs.\ $r_{\tau}$ for all three gauge groups with fixed $\zeta = 0.3$.
From these plots, we can already see that the susceptibility peak gets sharper as $N$ increases.
We make this statement more precise in \fig{Fig:scaling}, which shows $\chi_{\text{max}}$ vs.\ $N^2$ with log--log axes.
By fitting the power law \eq{eq:power}, we obtain $b = 0.61(8)$.
This suggests a continuous second-order phase transition in the two-dimensional $\cN = (2, 2)$ SYM theory, in contrast to the first-order transition of $\cN = (8, 8)$ SYM.
However, note that our uncertainty on $b$ is purely statistical and doesn't take into account potential systematic effects.
For instance, it is likely that the true $\chi_{\text{max}}$ lies in between the discrete $r_{\tau}$ points we have analyzed.
In the future, this can be improved by using multi-ensemble reweighting to interpolate between these points~\cite{Kuramashi:2020meg, Ferrenberg:1988yz}.

Our result for the critical exponent $b$ also comes from a single, relatively small lattice size, $N_x = N_{\tau} = 12$.
Recall from \secref{sec:Lattice_setup} that larger $(N_x, N_{\tau})$ simply reduces the lattice spacing at the transition, with the $a \to 0$ continuum limit corresponding to $N_x = \al N_{\tau} \to \infty$.
We have started to study discretization artifacts for gauge group SU(12) by analyzing larger lattice sizes up to $32 \X 32$ (Table~\ref{Tab:config}).
For aspect ratio $\al = 1$, our results for $\vev{|W^u|}$ agree within statistical uncertainties across these lattice sizes.
However, the rapid increase in computational costs with the number of colors, $\sim N^{7 / 2}$, has so far prevented us from repeating our full scaling analysis for these larger systems.

\begin{figure*}[htbp]
  \includegraphics[width=0.49\linewidth]{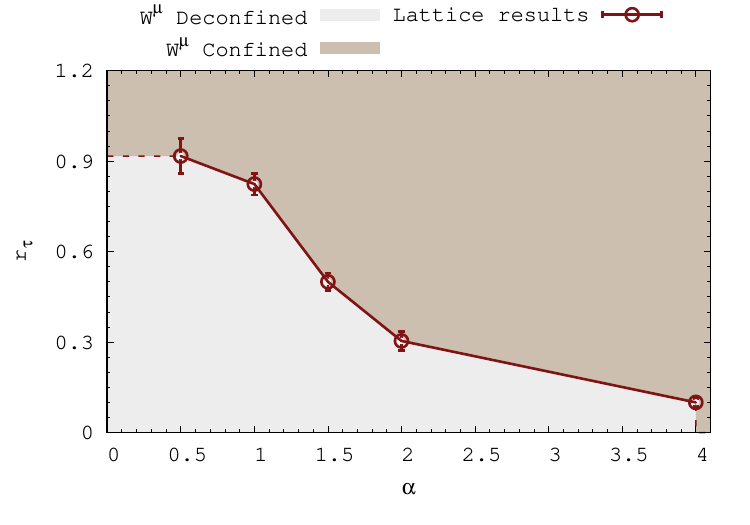} \hfill
  \includegraphics[width=0.49\linewidth]{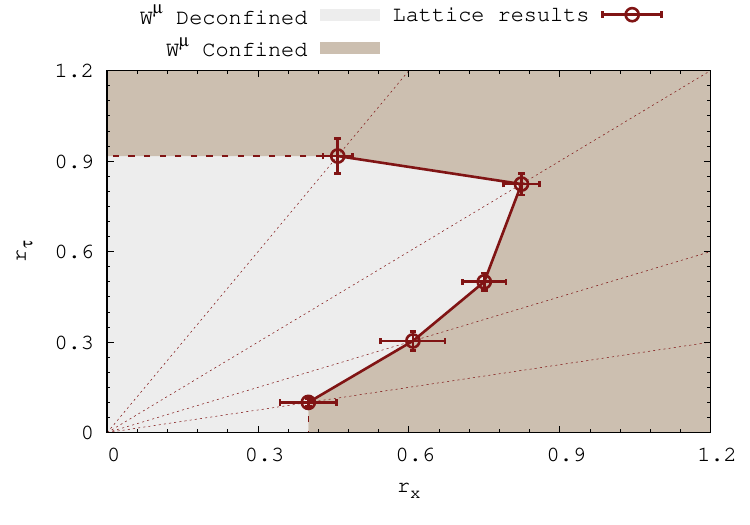}
  \caption{Two views of the phase diagram for two-dimensional $\cN = (2, 2)$ lattice SYM with gauge group SU(12), showing how the critical inverse temperature $r_{\tau}^{(c)}$ of the spatial deconfinement transition depends on the aspect ratio $\al = r_x / r_{\tau}$ (shown by dotted lines in the right plot) in the $\zeta \to 0$ limit.  In contrast to $\cN = (8, 8)$ SYM, the transition always occurs at relatively high temperatures corresponding to $r_{\tau}^{(c)} \lesssim 1$.}
  \label{Fig:alpha_vs_rt}
\end{figure*}

Instead, we have focused on further exploring the phase structure of two-dimensional $\cN = (2, 2)$ SYM by analyzing different aspect ratios $\al = 1/2$, $3/2$, $2$, and $4$, in each case for gauge group SU(12) and a single lattice size with $\max\{N_x, N_{\tau}\} = 24$ (Table~\ref{Tab:config}).
For each aspect ratio, we use the spatial Wilson line susceptibility to determine the critical $r_{\tau}^{(c)}$ that we present in \fig{Fig:alpha_vs_rt}.
For all of these results, we use three values of $\zeta$ to extrapolate to the $\zeta \to 0$ limit where the soft supersymmetry breaking from \eq{eq:soft} vanishes.

For large aspect ratios $\al \geq 3/2$ where the spatial deconfinement transition occurs at relatively high temperatures corresponding to $r_{\tau}^{(c)} \lesssim 0.5$, we find behavior similar to the two-dimensional $\cN = (8, 8)$ SYM case~\cite{Catterall:2017lub}.
In particular, the transition moves to lower temperatures (larger $r_{\tau}$) as \al decreases.
For $\cN = (8, 8)$ SYM, this behavior persists into the low-temperature holographic regime $r_{\tau} \gg 1$, where the transition is predicted by dual supergravity solutions.
Our new results in \fig{Fig:alpha_vs_rt} indicate that $\cN = (2, 2)$ SYM behaves differently: For smaller $\al \leq 1$, the critical inverse temperature becomes roughly constant, $r_{\tau}^{(c)} \approx 1$.

While this is a striking contrast compared to the maximally supersymmetric theory, the absence of a spatial deconfinement transition in the low-temperature region where holography would be valid, $r_\tau \gg 1$, does not rule out the existence of a gravity dual for two-dimensional $\cN = (2, 2)$ SYM.
Instead, it indicates that any low-temperature dual gravity solution would not undergo a topology-changing transition between a phase dominated by homogeneous D1 (black string) solutions wrapping around the spatial circle and a phase dominated by D0 (black hole) solutions localized on the spatial circle. 

\section{Conclusions}
\label{Sec:Conclusions}

In this work, we have used non-perturbative lattice field theory calculations to study the phase diagram of two-dimensional $\cN = (2, 2)$ SYM.
We first formulated the finite-temperature lattice theory using a lattice action that preserves one of the four supercharges at non-zero lattice spacing.
From numerical calculations, we observed spatial deconfinement transitions for a range of aspect ratios $1/2 \leq \al \leq 4$ and determined the critical inverse temperatures $r_{\tau}^{(c)}$.
For the case $\al = 1$, we additionally compared results for gauge groups SU(12), SU(16), and SU(20) to estimate that this may be a continuous second-order phase transition.

Although our high-temperature results corresponding to $\al \geq 3/2$ are qualitatively similar to those obtained for $\cN = (8, 8)$ SYM~\cite{Catterall:2017lub}, striking differences appear at lower temperatures (aspect ratios $\al \leq 1$).
In particular, the transition does not continue moving to lower temperatures for smaller aspect ratios.
Instead, the critical inverse temperature becomes roughly constant around $r_{\tau}^{(c)} \approx 1$.
That is, the transition does not persist in the low-temperature regime where any holographic gravity solution would be valid.
This suggests that the holographic dual of this gauge theory, if it exists, does not undergo a black hole topology-changing transition at low temperatures.

The results presented in this work lay the groundwork for further lattice investigations of $\cN = (2, 2)$ SYM at finite temperature.
In particular, we are analyzing the `extent of the scalars', $\Tr{X^2}$, again in order to compare this theory against $\cN = (8, 8)$ SYM, where large-$N$ holography relates this observable to bound states of D-branes in the dual gravitational solutions.
We recently presented preliminary results from this analysis in \refcite{Dhindsa:2021irw}.
Another target of our work is to investigate how the eigenvalues of the scalars spread among themselves, which is interesting because it would be holographically related to the `extent of space'.

We can also build on the work reported here by analyzing related lattice field theories.
For example, \refcite{Aharony:2005ew} provides an incomplete phase diagram of the two-dimensional bosonic theory obtained by removing the fermions from $\cN = (2, 2)$ SYM.
It would be interesting to reproduce and extend this phase diagram using non-perturbative lattice calculations.
Another interesting problem for the future will be to study two-dimensional $\cN = (4, 4)$ SYM.
This would complete the family of two-dimensional pure SYM theories for which lattice calculations can preserve a subset of the supersymmetries at non-zero lattice spacing.
Comparing all three theories with four, eight, and sixteen supercharges would test whether or not the continuation of the spatial deconfinement transition to the low-temperature holographic regime is a unique feature of the maximally supersymmetric case.

\begin{table*}[htpb]
  \renewcommand\arraystretch{1.2}  
  \addtolength{\tabcolsep}{3 pt}   
  \begin{tabular}{c|c|c|c|c|c}
    \hline
    Gauge group              & $\alpha$            & $N_x \X N_\tau$ &  Range of $r_\tau$  & $\zeta$       & \# ensembles \\
    \hline
    \multirowcell{8}{SU(12)} & $\frac{1}{2}$       & $12\X 24$       & 0.8 - 2.0           & 0.3, 0.4, 0.5 & 33           \\
    \cline{2-6}
                             & \multirowcell{4}{1} & $12\X 12$       & 0.6 - 5.0           & 0.2, 0.3, 0.4 & 48           \\
                             &                     & $16\X 16$       & 3.0 - 5.0           & 0.3, 0.4, 0.5 &  6           \\
                             &                     & $24\X 24$       & 0.333 - 3.0         & 0.3, 0.4, 0.5 & 21           \\
                             &                     & $32\X 32$       & 1.0 - 3.0           & 0.3, 0.4, 0.5 &  6           \\
    \cline{2-6}
                             & $\frac{3}{2}$       & $24\X 16$       & 0.3 - 0.9           & 0.3, 0.4, 0.5 & 39           \\
    \cline{2-6}
                             & 2                   & $24\X 12$       & 0.2 - 1.4           & 0.3, 0.4, 0.5 & 33           \\
    \cline{2-6}
                             & 4                   & $24\X 6$        & 0.05 - 0.2          & 0.6, 0.7, 0.8 & 21           \\
    \hline
    SU(16)                   & 1                   & $12\X 12$       & 0.6 - 1.6           & 0.3           & 12           \\
    \hline
    SU(20)                   & 1                   & $12\X 12$       & 0.6 - 1.6           & 0.3           & 12           \\
    \hline
  \end{tabular}
  \caption{Summary of the 231 lattice ensembles used in our numerical calculations, with full information available in \refcite{data}.}
  \label{Tab:config}
\end{table*}

\section*{Acknowledgements}
\label{Sec:Acknowledgement}

We thank Toby Wiseman for helpful discussions. This research was supported in part by the International Centre for Theoretical Sciences (ICTS) through the program ``Nonperturbative and Numerical Approaches to Quantum Gravity, String Theory and Holography'' (code: ICTS/numstrings-2022/8).
NSD thanks the Council of Scientific and Industrial Research (CSIR), Government of India, for the financial support through a research fellowship (Award No.~{09/947(0119)/2019-EMR-I}) during the initial phases of this work.
RGJ was supported by the U.S. Department of Energy (DOE), Office of Science, National Quantum Information Science Research Centers, Co-design Center for Quantum Advantage under contract No.~{DE-SC0012704}. RGJ was also supported by the DOE with contract No.~{DE-AC05-06OR23177}, under which Jefferson Science Associates, LLC operates Jefferson Lab.
The work of AJ was supported in part by the Start-up Research Grant from the University of the Witwatersrand.
DS was supported by UK Research and Innovation Future Leader Fellowship {MR/S015418/1} \& {MR/X015157/1} and STFC grants {ST/T000988/1} \& {ST/X000699/1}.
Numerical calculations were carried out at the University of Liverpool and NABI, Mohali. We acknowledge the National Supercomputing Mission (NSM) for providing computing resources of PARAM SMRITI at NABI, Mohali, which is implemented by C-DAC and supported by the Ministry of Electronics and Information Technology (MeitY) and the Department of Science and Technology (DST), Government of India. \\[8 pt]

\noindent \textbf{Data Availability Statement:} All data used in this work are available through the open data release \refcite{data}, which also provides the bulk of the computational workflow needed to reproduce, check, and extend our analyses.

\appendix
\section{Summary of lattice ensembles}
\label{Sec:Lattice_ensembles}

Table~\ref{Tab:config} summarizes the 231 lattice ensembles used in our numerical analyses.
As discussed in \secref{sec:Lattice_calcs}, we consider a range of aspect ratios $1/2 \leq \al \leq 4$, and for $\al = 1$ additionally investigate several lattice sizes up to $32\times 32$, and three SU($N$) gauge groups up to $N = 20$.
For each ensemble we typically generate 3500 molecular dynamics time units (MDTU), using unit-length trajectories in the RHMC algorithm, and impose a thermalization cut after the first 500~MDTU.
Simple observables including the Maldacena loop, unitarized Wilson line, and the extent of the scalars $\Tr{X^2}$ are measured after every trajectory.
More involved measurements including the stochastic computation of the fermion bilinear term that appears in the \cQ Ward identity \eq{eq:bilin} are done using configurations saved to disk every 10~MDTU.
Another measurement done using these saved configurations is to compute the extremal eigenvalues of the squared fermion operator $\Ddag D$, where $\Psi^T D \Psi$ corresponds to the terms in the action \eq{eq:exact} involving the fermions $\Psi^T = (\eta, \psi_a, \chi_{ab})$.
These eigenvalue computations are performed using a Davidson-type method provided by the PReconditioned Iterative Multi-Method Eigensolver (PRIMME) library~\cite{Stathopoulos:2010}.

After generating configurations and carrying out these measurements, we use the `autocorr' module in \texttt{emcee}~\cite{Foreman:2013mc} to estimate auto-correlation times $\tau$ for four relevant quantities.
These are the magnitude of the Maldacena loop $|M|$, the lowest eigenvalue of $\Ddag D$, the Ward identity \eq{eq:bilin} that involves the $\eta\psi$ fermion bilinear, and $\Tr{X^2}$.
The resulting auto-correlation times confirm a posteriori that a minimum of 5$\tau$, and typically well over 10$\tau$, are set aside for thermalization.
We divide our thermalized measurements into blocks for jackknife analyses, with block sizes larger than $\tau$, and at least 30~MDTU.
Our approach ensures that at least 27 such statistically independent blocks are available for each ensemble, more than enough for robust analyses.
This information is summarized in Tables~\ref{tab:12nt24}--\ref{tab:largeN}, which are extracted from the open data release \refcite{data}.
That data release also provides a great deal of additional information including extremal eigenvalues of $\Ddag D$ (which must remain within the spectral range where the rational approximation used in the RHMC algorithm is reliable), and other observables computed in addition to the spatial Wilson line and Polyakov loop discussed above.

\begin{table*}[htpb]
  \renewcommand\arraystretch{1.2}  
  \addtolength{\tabcolsep}{3 pt}   
  \begin{tabular}{l|c|r|r|r|r|r|r}
    \hline
    $r_{\tau}$ & $\zeta$ & $\tau_{|M|}$ & $\tau_{\text{eig}}$ & $\tau_{\eta\psi}$ & $\tau_{X^2}$ & Block & Num. \\
    \hline
               & 0.3     & 22           &  9                  & 19                & 10           & 30    & 100  \\
    0.8        & 0.4     & 16           & 11                  & 11                & 17           & 30    & 100  \\
               & 0.5     & 18           &  9                  &  6                & 20           & 30    & 100  \\
    \hline
               & 0.3     & 31           & 12                  & 11                & 19           & 40    &  75  \\
    0.9        & 0.4     & 14           & 10                  & 14                & 25           & 30    & 100  \\
               & 0.5     & 13           &  9                  & 10                & 19           & 30    & 100  \\
    \hline
               & 0.3     & 30           &  8                  & 10                & 14           & 40    &  75  \\
    0.95       & 0.4     & 18           & 11                  &  7                & 30           & 40    &  75  \\
               & 0.5     & 14           & 11                  & 11                & 13           & 30    & 100  \\
    \hline
               & 0.3     & 25           &  8                  & 12                & 17           & 30    & 100  \\
    1.0        & 0.4     & 12           & 11                  & 17                & 15           & 30    & 100  \\
               & 0.5     & 12           & 15                  &  9                & 17           & 30    & 100  \\
    \hline
               & 0.3     & 15           & 15                  &  6                & 14           & 30    & 100  \\
    1.05       & 0.4     & 22           & 15                  & 18                & 18           & 30    & 100  \\
               & 0.5     &  9           &  7                  & 17                & 15           & 30    & 100  \\
    \hline
               & 0.3     & 19           & 13                  &  7                & 14           & 30    & 100  \\
    1.1        & 0.4     & 17           & 13                  & 13                & 14           & 30    & 100  \\
               & 0.5     & 10           & 22                  &  7                & 12           & 30    & 100  \\
    \hline
               & 0.3     & 29           &  8                  & 11                & 10           & 30    & 100  \\
    1.2        & 0.4     & 12           &  9                  &  9                &  7           & 30    & 100  \\
               & 0.5     &  8           & 10                  &  9                &  9           & 30    & 100  \\
    \hline
               & 0.3     & 20           & 13                  & 10                &  9           & 30    & 100  \\
    1.4        & 0.4     & 18           &  9                  &  9                & 10           & 30    & 100  \\
               & 0.5     &  7           &  6                  & 12                & 15           & 30    & 100  \\
    \hline
               & 0.3     & 10           & 12                  &  7                & 15           & 30    & 100  \\
    1.6        & 0.4     &  9           & 10                  &  8                & 12           & 30    & 100  \\
               & 0.5     &  8           & 11                  &  5                &  7           & 30    & 100  \\
    \hline
               & 0.3     & 16           & 10                  & 11                & 12           & 30    & 100  \\
    1.8        & 0.4     & 11           & 11                  & 10                & 13           & 30    & 100  \\
               & 0.5     & 17           & 11                  & 19                & 12           & 30    & 100  \\
    \hline
               & 0.3     & 10           &  8                  &  9                & 15           & 30    & 100  \\
    2.0        & 0.4     & 16           & 13                  & 10                & 10           & 30    & 100  \\
               & 0.5     &  9           & 15                  & 13                &  7           & 30    & 100  \\
    \hline
  \end{tabular}
  \caption{\label{tab:12nt24}Summary of the thermalized data available for the 33 lattice ensembles with gauge group SU(12) and lattice size $12\X 24$ (aspect ratio $\al = \frac{1}{2}$), extracted from the full information available in \refcite{data}.  The four columns labelled $\tau$ report integer auto-correlation times (in MDTU) for the Maldacena loop magnitude ($|M|$), the lowest $\Ddag D$ eigenvalue (eig), the Ward identity involving the $\eta\psi$ fermion bilinear, and the extent of the scalars $\Tr{X^2}$.  The final two columns are the MDTU block size used in jackknife analyses, and the resulting number of jackknife blocks following thermalization.}
\end{table*}

\begin{table*}[htpb]
  \renewcommand\arraystretch{1.2}  
  \addtolength{\tabcolsep}{3 pt}   
  \begin{tabular}{l|c|r|r|r|r|r|r}
    \hline
    $r_{\tau}$ & $\zeta$ & $\tau_{|M|}$ & $\tau_{\text{eig}}$ & $\tau_{\eta\psi}$ & $\tau_{X^2}$ & Block & Num. \\
    \hline
               & 0.2     &  51          & 15                  &  9                & 25           &  60    &  50 \\
    0.6        & 0.3     &  22          &  7                  &  4                & 11           &  30    & 100 \\
               & 0.4     &  14          & 10                  &  9                & 13           &  30    & 100 \\
    \hline
               & 0.2     &  24          & 10                  & 17                & 24           &  30    & 100 \\
    0.7        & 0.3     &  12          & 13                  & 11                &  9           &  30    & 100 \\
               & 0.4     &   9          & 10                  &  6                & 17           &  30    & 100 \\
    \hline
               & 0.2     &  29          & 24                  & 13                & 27           &  30    & 100 \\
    0.75       & 0.3     &  20          & 13                  & 12                & 13           &  30    & 100 \\
               & 0.4     &  11          & 11                  & 11                & 11           &  30    & 100 \\
    \hline
               & 0.2     &  37          & 12                  & 10                & 26           &  40    &  75 \\
    0.8        & 0.3     &  17          & 11                  &  7                & 10           &  30    & 100 \\
               & 0.4     &  14          &  5                  & 12                & 11           &  30    & 100 \\
    \hline
               & 0.2     &  21          & 15                  & 12                & 16           &  30    & 100 \\
    0.85       & 0.3     &  18          &  7                  & 11                & 10           &  30    & 100 \\
               & 0.4     &  10          &  7                  &  5                &  9           &  30    & 100 \\
    \hline
               & 0.2     &  33          & 14                  &  8                & 19           &  40    &  75 \\
    0.9        & 0.3     &  23          & 20                  &  9                & 19           &  30    & 100 \\
               & 0.4     &  14          & 13                  &  8                & 14           &  30    & 100 \\
    \hline
               & 0.2     & 100          & 17                  &  8                & 27           & 110    &  27 \\
    0.95       & 0.3     &  18          &  8                  & 12                & 13           &  30    & 100 \\
               & 0.4     &  13          & 17                  &  7                & 10           &  30    & 100 \\
    \hline
               & 0.2     &  59          &  9                  & 15                & 58           &  60    &  50 \\
    1.0        & 0.3     &  17          &  9                  & 11                & 17           &  30    & 100 \\
               & 0.4     &  22          &  9                  & 14                & 10           &  30    & 100 \\
    \hline
               & 0.2     &  25          & 13                  &  9                & 28           &  30    & 100 \\
    1.1        & 0.3     &  18          & 15                  &  9                & 16           &  30    & 100 \\
               & 0.4     &  11          &  8                  &  5                & 11           &  30    & 100 \\
    \hline
               & 0.2     &  35          & 22                  &  8                & 34           &  40    &  75 \\
    1.2        & 0.3     &  19          & 12                  &  7                & 14           &  30    & 100 \\
               & 0.4     &  16          & 12                  & 11                &  7           &  30    & 100 \\
    \hline
               & 0.2     &  47          & 13                  &  6                & 14           &  50    &  60 \\
    1.3        & 0.3     &  30          & 10                  &  6                & 11           &  40    &  75 \\
               & 0.4     &  15          &  9                  &  8                & 14           &  30    & 100 \\
    \hline
               & 0.2     &  65          & 14                  &  7                & 18           &  70    &  42 \\
    1.6        & 0.3     &  21          &  8                  & 12                & 33           &  40    &  75 \\
               & 0.4     &  14          & 11                  &  9                & 14           &  30    & 100 \\
    \hline
               & 0.2     &  27          & 11                  &  9                & 31           &  40    &  75 \\
    2.0        & 0.3     &  17          & 15                  &  8                & 27           &  30    & 100 \\
               & 0.4     &   7          &  9                  &  8                & 12           &  30    & 100 \\
    \hline
               & 0.2     &  20          & 15                  & 10                & 35           &  40    &  75 \\
    3.0        & 0.3     &   7          &  7                  &  9                & 20           &  30    & 100 \\
               & 0.4     &   5          &  7                  &  9                & 14           &  30    & 100 \\
    \hline
               & 0.2     &  16          & 13                  &  9                & 40           &  50    &  59 \\
    4.0        & 0.3     &   7          &  7                  &  7                & 36           &  40    &  75 \\
               & 0.4     &   6          &  9                  &  7                & 22           &  30    & 100 \\
    \hline
               & 0.2     &   7          &  9                  &  8                & 16           &  30    & 100 \\
    5.0        & 0.3     &   4          & 12                  & 14                & 17           &  30    &  80 \\
               & 0.4     &   3          & 11                  &  4                & 22           &  30    &  48 \\
    \hline
  \end{tabular}
  \caption{\label{tab:12nt12}Summary of the thermalized data available for the 48 lattice ensembles with gauge group SU(12) and lattice size $12\X 12$ (aspect ratio $\al = 1$), extracted from the full information available in \refcite{data}, with columns as in \tab{tab:12nt24}.}
\end{table*}

\begin{table*}[htpb]
  \renewcommand\arraystretch{1.2}  
  \addtolength{\tabcolsep}{3 pt}   
  \begin{tabular}{c|l|c|r|r|r|r|r|r}
    \hline
    $N_x \X N_{\tau}$ & $r_{\tau}$ & $\zeta$ & $\tau_{|M|}$ & $\tau_{\text{eig}}$ & $\tau_{\eta\psi}$ & $\tau_{X^2}$ & Block & Num. \\
    \hline
                      &            & 0.3     & 12           &  9                  &  9                & 34           & 40    &  75  \\
    $16\X 16$         & 3.0        & 0.4     &  6           &  9                  &  7                & 15           & 30    & 100  \\
                      &            & 0.5     &  5           &  9                  & 10                & 14           & 30    & 100  \\
    \hline
                      &            & 0.3     &  5           &  9                  &  7                & 35           & 40    &  75  \\
    $16\X 16$         & 5.0        & 0.4     &  4           &  7                  & 11                & 26           & 30    &  98  \\
                      &            & 0.5     &  3           &  9                  &  8                & 28           & 30    & 100  \\
    \hline
                      &            & 0.3     & 18           &  7                  & 11                & 17           & 30    & 100  \\
    $24\X 24$         & 0.333      & 0.4     & 19           &  5                  & 12                & 17           & 30    & 100  \\
                      &            & 0.5     & 11           & 15                  &  5                & 8            & 30    & 100  \\
    \hline
                      &            & 0.3     & 20           & 11                  &  9                & 20           & 30    & 100  \\
    $24\X 24$         & 0.5        & 0.4     & 23           & 13                  & 12                & 13           & 30    & 100  \\
                      &            & 0.5     & 11           &  6                  &  9                & 9            & 30    & 100  \\
    \hline
                      &            & 0.3     & 41           & 11                  & 11                & 13           & 50    &  60  \\
    $24\X 24$         & 0.75       & 0.4     & 12           & 13                  & 11                & 14           & 30    & 100  \\
                      &            & 0.5     &  8           & 10                  &  6                & 11           & 30    & 100  \\
    \hline
                      &            & 0.3     & 30           & 13                  &  8                & 13           & 40    &  75  \\
    $24\X 24$         & 1.0        & 0.4     & 24           & 15                  & 18                & 11           & 30    & 100  \\
                      &            & 0.5     & 12           & 11                  & 10                & 15           & 30    & 100  \\
    \hline
                      &            & 0.3     & 41           &  9                  &  7                & 14           & 50    &  60  \\
    $24\X 24$         & 1.25       & 0.4     & 17           &  7                  &  8                & 8            & 30    & 100  \\
                      &            & 0.5     & 10           &  7                  &  9                & 8            & 30    & 100  \\
    \hline
                      &            & 0.3     & 15           &  9                  &  8                & 11           & 30    & 100  \\
    $24\X 24$         & 2.0        & 0.4     & 16           &  7                  & 10                & 30           & 40    &  75  \\
                      &            & 0.5     &  5           & 10                  &  9                & 9            & 30    & 100  \\
    \hline
                      &            & 0.3     & 13           &  8                  &  4                & 34           & 40    &  75  \\
    $24\X 24$         & 3.0        & 0.4     &  8           &  7                  & 13                & 15           & 30    & 100  \\
                      &            & 0.5     &  5           &  7                  & 12                & 19           & 30    & 100  \\
    \hline
                      &            & 0.3     & 28           &  7                  &  8                & 19           & 30    & 100  \\
    $32\X 32$         & 1.0        & 0.4     & 10           & 10                  & 11                & 11           & 30    & 100  \\
                      &            & 0.5     & 20           & 12                  &  6                & 13           & 30    & 100  \\
    \hline
                      &            & 0.3     & 11           & 13                  &  9                & 25           & 30    & 100  \\
    $32\X 32$         & 3.0        & 0.4     &  7           &  9                  &  7                & 20           & 30    & 100  \\
                      &            & 0.5     &  6           &  6                  & 11                & 23           & 30    & 100  \\
    \hline
  \end{tabular}
  \caption{\label{tab:vol}Summary of the thermalized data available for the 33 lattice ensembles with gauge group SU(12) and larger lattice sizes $16\X 16$, $24\X 24$ and $32\X 32$ (aspect ratio $\al = 1$), extracted from the full information available in \refcite{data}, with columns as in \tab{tab:12nt24}.}
\end{table*}

\begin{table*}[htpb]
  \renewcommand\arraystretch{1.2}  
  \addtolength{\tabcolsep}{3 pt}   
  \begin{tabular}{l|c|r|r|r|r|r|r}
    \hline
    $r_{\tau}$ & $\zeta$ & $\tau_{|M|}$ & $\tau_{\text{eig}}$ & $\tau_{\eta\psi}$ & $\tau_{X^2}$ & Block & Num. \\
    \hline
               & 0.3     & 28           & 10                  &  5            & 15               & 30    & 100  \\
    0.3        & 0.4     & 13           & 11                  & 11            &  9               & 30    & 100  \\
               & 0.5     & 14           &  8                  & 11            & 10               & 30    & 100  \\
    \hline
               & 0.3     & 31           & 13                  &  8            & 27               & 40    &  75  \\
    0.35       & 0.4     & 12           &  7                  &  7            & 13               & 30    & 100  \\
               & 0.5     & 10           &  7                  & 16            & 10               & 30    & 100  \\
    \hline
               & 0.3     & 18           & 11                  & 12            & 18               & 30    & 100  \\
    0.4        & 0.4     & 13           & 10                  &  5            & 10               & 30    & 100  \\
               & 0.5     & 11           & 11                  &  7            &  7               & 30    & 100  \\
    \hline
               & 0.3     & 16           & 16                  & 14            &  9               & 30    & 100  \\
    0.45       & 0.4     & 19           & 12                  &  7            & 11               & 30    & 100  \\
               & 0.5     &  8           & 13                  &  7            &  9               & 30    & 100  \\
    \hline
               & 0.3     & 17           & 19                  &  9            & 14               & 30    & 100  \\
    0.5        & 0.4     & 19           & 13                  & 12            & 11               & 30    & 100  \\
               & 0.5     & 12           & 17                  &  7            & 13               & 30    & 100  \\
    \hline
               & 0.3     & 34           & 21                  &  9            & 17               & 40    &  75  \\
    0.55       & 0.4     & 14           &  9                  &  9            &  7               & 30    & 100  \\
               & 0.5     & 14           & 11                  &  7            &  8               & 30    & 100  \\
    \hline
               & 0.3     & 22           & 13                  &  7            &  8               & 30    & 100  \\
    0.6        & 0.4     & 17           &  9                  & 11            &  8               & 30    & 100  \\
               & 0.5     & 15           & 12                  & 16            & 11               & 30    & 100  \\
    \hline
               & 0.3     & 28           &  8                  &  6            &  9               & 30    & 100  \\
    0.65       & 0.4     & 21           &  9                  &  6            &  6               & 30    & 100  \\
               & 0.5     &  7           &  9                  &  9            &  8               & 30    & 100  \\
    \hline
               & 0.3     & 30           &  7                  & 15            & 17               & 40    &  75  \\
    0.7        & 0.4     & 11           &  9                  & 14            & 17               & 30    & 100  \\
               & 0.5     &  7           &  7                  &  9            & 11               & 30    & 100  \\
    \hline
               & 0.3     & 49           &  8                  &  6            & 12               & 50    &  60  \\
    0.75       & 0.4     &  9           &  7                  &  8            &  9               & 30    & 100  \\
               & 0.5     &  8           &  8                  &  9            & 11               & 30    & 100  \\
    \hline
               & 0.3     & 17           & 11                  &  9            &  7               & 30    & 100  \\
    0.8        & 0.4     & 10           & 13                  & 12            &  9               & 30    & 100  \\
               & 0.5     &  8           &  9                  & 14            & 15               & 30    & 100  \\
    \hline
               & 0.3     & 23           & 11                  &  6            & 10               & 30    & 100  \\
    0.85       & 0.4     & 14           & 12                  & 15            &  9               & 30    & 100  \\
               & 0.5     &  9           &  8                  &  6            & 14               & 30    & 100  \\
    \hline
               & 0.3     & 18           & 12                  & 10            & 13               & 30    & 100  \\
    0.9        & 0.4     & 16           &  7                  & 13            & 15               & 30    & 100  \\
               & 0.5     &  9           & 14                  &  8            &  7               & 30    & 100  \\
    \hline
  \end{tabular}
  \caption{\label{tab:24nt16}Summary of the thermalized data available for the 39 lattice ensembles with gauge group SU(12) and lattice size $24\X 16$ (aspect ratio $\al = \frac{3}{2}$), extracted from the full information available in \refcite{data}, with columns as in \tab{tab:12nt24}.}
\end{table*}

\begin{table*}[htpb]
  \renewcommand\arraystretch{1.2}  
  \addtolength{\tabcolsep}{3 pt}   
  \begin{tabular}{l|c|r|r|r|r|r|r}
    \hline
    $r_{\tau}$ & $\zeta$ & $\tau_{|M|}$ & $\tau_{\text{eig}}$ & $\tau_{\eta\psi}$ & $\tau_{X^2}$ & Block & Num. \\
    \hline
               & 0.3     & 14           & 11                  & 19                & 21           & 30    & 100  \\
    0.2        & 0.4     & 10           &  8                  & 10                & 7            & 30    & 100  \\
               & 0.5     & 17           & 12                  & 10                & 19           & 30    & 100  \\
    \hline
               & 0.3     & 27           & 15                  &  9                & 39           & 40    &  75  \\
    0.25       & 0.4     &  8           & 12                  &  7                & 11           & 30    & 100  \\
               & 0.5     &  8           &  8                  &  9                & 9            & 30    & 100  \\
    \hline
               & 0.3     & 16           & 10                  & 12                & 15           & 30    & 100  \\
    0.3        & 0.4     & 20           & 15                  & 13                & 14           & 30    & 100  \\
               & 0.5     & 13           & 21                  &  9                & 9            & 30    & 100  \\
    \hline
               & 0.3     & 21           & 10                  & 14                & 28           & 30    & 100  \\
    0.35       & 0.4     &  9           & 10                  &  7                & 7            & 30    & 100  \\
               & 0.5     &  7           & 10                  &  9                & 10           & 30    & 100  \\
    \hline
               & 0.3     & 56           & 18                  &  8                & 26           & 60    &  50  \\
    0.4        & 0.4     & 21           &  7                  &  9                & 12           & 30    & 100  \\
               & 0.5     & 15           &  9                  &  8                & 7            & 30    & 100  \\
    \hline
               & 0.3     & 31           &  9                  &  7                & 33           & 40    &  75  \\
    0.5        & 0.4     & 14           &  9                  &  5                & 12           & 30    & 100  \\
               & 0.5     & 14           &  7                  & 11                & 7            & 30    & 100  \\
    \hline
               & 0.3     & 20           & 13                  & 11                & 13           & 30    & 100  \\
    0.6        & 0.4     & 12           &  8                  &  7                & 11           & 30    & 100  \\
               & 0.5     & 10           & 10                  &  7                & 11           & 30    & 100  \\
    \hline
               & 0.3     & 34           &  6                  &  9                & 14           & 40    &  75  \\
    0.8        & 0.4     &  9           &  7                  & 13                & 12           & 30    & 100  \\
               & 0.5     &  8           &  7                  &  9                & 10           & 30    & 100  \\
    \hline
               & 0.3     & 31           & 10                  & 11                & 13           & 40    &  75  \\
    1.0        & 0.4     &  7           &  7                  &  8                & 7            & 30    & 100  \\
               & 0.5     &  8           &  9                  &  7                & 10           & 30    & 100  \\
    \hline
               & 0.3     & 15           & 11                  & 15                & 18           & 30    & 100  \\
    1.2        & 0.4     & 11           & 13                  &  5                & 11           & 30    & 100  \\
               & 0.5     &  6           &  7                  & 10                & 10           & 30    & 100  \\
    \hline
               & 0.3     & 16           &  8                  & 10                & 12           & 30    & 100  \\
    1.4        & 0.4     &  8           &  9                  &  9                & 12           & 30    & 100  \\
               & 0.5     &  6           &  5                  & 11                & 9            & 30    & 100  \\
    \hline
  \end{tabular}
  \caption{\label{tab:24nt12}Summary of the thermalized data available for the 33 lattice ensembles with gauge group SU(12) and lattice size $24\X 12$ (aspect ratio $\al = 2$), extracted from the full information available in \refcite{data}, with columns as in \tab{tab:12nt24}.}
\end{table*}

\begin{table*}[htpb]
  \renewcommand\arraystretch{1.2}  
  \addtolength{\tabcolsep}{3 pt}   
  \begin{tabular}{l|c|r|r|r|r}
    \hline
    $r_{\tau}$ & $\zeta$ & $\tau_{|M|}$ & $\tau_{\text{eig}}$ & $\tau_{\eta\psi}$ & $\tau_{X^2}$ \\
    \hline
               & 0.6     & 6            &  8                  & 14                &  8           \\
    0.05       & 0.7     & 6            &  9                  & 11                & 11           \\
               & 0.8     & 6            &  9                  & 13                & 17           \\
    \hline
               & 0.6     & 6            &  5                  &  8                & 11           \\
    0.075      & 0.7     & 5            & 11                  &  7                &  6           \\
               & 0.8     & 5            &  9                  &  5                & 11           \\
    \hline
               & 0.6     & 7            &  7                  &  9                & 12           \\
    0.1        & 0.7     & 5            & 11                  & 11                &  8           \\
               & 0.8     & 5            & 12                  & 10                &  7           \\
    \hline
               & 0.6     & 4            &  8                  & 11                & 10           \\
    0.125      & 0.7     & 6            &  7                  &  9                &  5           \\
               & 0.8     & 6            &  5                  & 13                & 11           \\
    \hline
               & 0.6     & 6            &  7                  & 11                & 16           \\
    0.15       & 0.7     & 5            &  8                  &  7                & 10           \\
               & 0.8     & 6            &  5                  &  6                &  7           \\
    \hline
               & 0.6     & 7            &  9                  &  9                & 14           \\
    0.175      & 0.7     & 8            &  6                  &  7                &  8           \\
               & 0.8     & 8            &  8                  &  9                & 10           \\
    \hline
               & 0.6     & 7            &  8                  & 10                & 11           \\
    0.2        & 0.7     & 6            &  9                  & 14                &  9           \\
               & 0.8     & 9            &  9                  &  7                &  8           \\
    \hline
  \end{tabular}
  \caption{\label{tab:24nt6}Summary of the thermalized data available for the 21 lattice ensembles with gauge group SU(12) and lattice size $24\X 6$ (aspect ratio $\al = 4$), extracted from the full information available in \refcite{data}.  The columns are as in \tab{tab:12nt24}, omitting the constant 30-MDTU block size and resulting 100 blocks.}
\end{table*}

\begin{table*}[htpb]
  \renewcommand\arraystretch{1.2}  
  \addtolength{\tabcolsep}{3 pt}   
  \begin{tabular}{c|l|r|r|r|r|r|r}
    \hline
    $N$ & $r_{\tau}$ & $\tau_{|M|}$ & $\tau_{\text{eig}}$ & $\tau_{\eta\psi}$ & $\tau_{X^2}$ & Block & Num. \\
    \hline
        & 0.6        & 3            & 12                  &  8                & 17           &  40   &  75  \\
        & 0.7        & 34           &  9                  & 11                & 29           &  40   &  75  \\
        & 0.75       & 25           & 11                  & 11                & 15           &  30   & 100  \\
        & 0.8        & 34           & 13                  &  7                & 10           &  40   &  75  \\
        & 0.85       & 23           & 11                  &  7                &  8           &  30   & 100  \\
    16  & 0.9        & 13           &  7                  & 11                &  7           &  30   & 100  \\
        & 0.95       & 36           & 11                  &  7                & 14           &  40   &  75  \\
        & 1.0        & 27           & 15                  &  9                & 11           &  30   & 100  \\
        & 1.1        & 18           &  8                  & 10                & 17           &  30   & 100  \\
        & 1.2        & 15           & 11                  &  9                & 17           &  30   & 100  \\
        & 1.3        & 27           & 13                  &  6                & 13           &  30   & 100  \\
        & 1.6        & 25           &  7                  &  9                & 31           &  40   &  75  \\
    \hline
        & 0.6        & 22           & 19                  & 18                & 15           &  30   & 100  \\
        & 0.7        & 33           & 16                  &  9                & 15           &  40   &  75  \\
        & 0.75       & 28           & 11                  & 10                &  7           &  30   & 100  \\
        & 0.8        & 26           & 10                  & 11                & 11           &  30   & 100  \\
        & 0.85       & 99           & 13                  &  9                & 10           & 100   &  30  \\
    20  & 0.9        & 33           & 13                  &  9                & 19           &  40   &  75  \\
        & 0.95       & 25           &  9                  &  8                & 35           &  40   &  75  \\
        & 1.0        & 43           &  8                  &  6                & 14           &  50   &  60  \\
        & 1.1        & 20           &  8                  & 11                & 10           &  30   & 100  \\
        & 1.2        & 23           &  9                  &  8                & 11           &  30   & 100  \\
        & 1.3        & 48           & 13                  & 11                & 23           &  50   &  60  \\
        & 1.6        & 29           &  7                  &  9                & 43           &  50   &  60  \\
    \hline
  \end{tabular}
  \caption{\label{tab:largeN}Summary of the thermalized data available for the 24 lattice ensembles with gauge group SU($N$) and lattice size $12\X 12$ (aspect ratio $\al = 1$), extracted from the full information available in \refcite{data}.  The columns are as in \tab{tab:12nt24}, omitting the constant $\zeta  = 0.3$.}
\end{table*}

\raggedright
\bibliography{v1}
\end{document}